\documentclass[fleqn,12pt]{article}

\newcommand{\AmS}{{\protect\the\textfont2
  A\kern-.1667em\lower.5ex\hbox{M}\kern-.125emS}}

\title{Can Radiogenic Heat Sources Inside the Earth be located by their Antineutrino incoming Directions?}

\author{G. Domogatsky$^1$, V. Kopeikin$^2$, L. Mikaelyan$^2$, V. Sinev$^2$ \\
\\
$^{1}$Institute for Nuclear Research RAS, Moscow, \\
$^{2}$Russian Research Center ``Kurchatov Institute"}
\date{} 

\begin{document}

\maketitle


\begin{abstract}
Antineutrinos born in the U and Th decay chains inside the Earth (``Geoneutrinos'') carry out information on the amount and
distribution of radiogenic heat sources, which is of fundamental importance for geophysics. Models of the Earth distribute U
and Th masses mainly between the continental crust and the lower mantle. It has been much discussed recently that a number
of detectors stationed at appropriate geographical sites can separate the crust and mantle contributions. In present work we
analyze directional separation of $\bar{{\nu}_e}$ signals arriving from the crust and the lower mantle with only one detector.
We find that with a $\sim$30-kton liquid scintillation antineutrino spectrometer using \, $\bar{{\nu}_e}+p \rightarrow n + e^{+}$
detection reaction and positron and neutron coordinates reconstruction techniques the U and Th distribution model can
roughly be tested. We also consider detector calibration using $\sim$1 MCi commercially available $^{90}$Sr-$^{90}$Y
beta source which emits $\bar{{\nu}_e}$ ($E_{\nu max}$ = 2.28 MeV) in the geoneutrino energy range.
\end{abstract}

\section*{Introduction}

The Earth emanates about 40 TW of heat flow coming from the interior [1]. It is believed that considerable part of this heat
originates from radioactive decay of Uranium, Thorium and Potassium hidden in the Earth. Antineutrinos $\bar{{\nu}_e}$ born
in the radioactive decays (``Geoneutrinos'') bring information on the abundances and radiogenic heat sources inside the Earth,
which are of a key importance for understanding of the formation and subsequent evolution of our planet.

The Geoneutrino concept is now 45 years old. A short overview of this concept can be found in our paper [2]. Only recently it
was recognized that Geoneutrinos emitted in the Uranium and Thorium decay chains can be detected in large volume liquid
scintillation underground spectrometers [3,4] using the inverse beta decay process

\begin{equation}
\bar{{\nu}_e}+p \rightarrow n + e^{+}
\end{equation}
as detection reaction. Actually KamLAND Collaboration in Japan has already demonstrated revolutionary progress in
$\bar{{\nu}_e}$ detection technique and has reported that a few geoneutrino events have been observed [5]. BOREXINO
Collaboration at LNGS, Italy, is planning experiments in the same field.

U and Th abundances and their distributions in the Earth are not known, except for a thin layer near the surface where direct
sampling is possible. Some information is obtained from analysis of volcanic outflows. Orthodox model of the Earth (sometimes
labeled as BSE model) and geochemichal arguments distribute U and Th masses mainly between the continental crust and the
lower mantle and thus the geoneutrino flux is expected to vary with geographical position of the observational point, decreasing
from continental to oceanic sites (Table 1). 

\begin{table}[htb]
\caption{Expected rates of U + Th geoneutrino events at various sites (exposition $10^{32}\;{\rm H\cdot year ^*}$,
no oscillations)}
\label{table}
\vspace{3pt}
\hspace{10pt}
\begin{tabular}{|c|ccc|}
\hline
Site & Rothschild {\it et al.$^{**}$} & Raghavan {\it et al.} & Fiorentini {\it et al.}\\
& [3] & [4] & [6,7]\\
\hline
Himalaya & 65 & - & 112 \\
Baksan & - & - & 91 \\
Gran Sasso & 53 & Ia: 134; Ib: 50 & 71 \\
Kamioka & 48 & Ia: 75; Ib: 50 & 61 \\
Hawaii & 27 & - & 22 \\
\hline
\end{tabular}\\[2pt]
{\small $^{*}$ 1160 ton of $\rm CH_2$ contain $10^{32}$ hydrogen atoms.}\\
{\small $^{**}$ Calculated using $\bar{{\nu}_e}$ fluxes from Ref. [3].}
\end{table}

As can be seen in Table 1, at continental sites dominate geoneutrinos born in the crust, at oceanic sites antineutrino flux is
a few times lower.

Geoneutrino data from several experiments around the world can separate crust and mantle contributions [3,4,6,7], and
thus to confirm or discard predictions of the Earth's model. Experiment of this type has been proposed for Baksan Neutrino
Observatory in North Caucasus [2].

The CHOOZ experiment in 1999 y demonstrated that the use of reaction (1) to detect low energy ($E<9$ MeV) reactor
 $\bar{{\nu}_e}$ in liquid scintillation detector provides determination of $\bar{{\nu}_e}$ incoming direction in case of
(practically) one point like $\bar{{\nu}_e}$ source [8]. 

In this paper we study separation capability of geoneutrino ($E<3.27$ MeV) signals coming from different directions (from
the crust and the mantle) using one $\sim$30-kton target mass liquid scintillation detector. At Baksan sample of $\sim$4000
geoneutrino events can be accumulated in $\sim$5 year (live time) data taking. We conclude with detector calibration procedure
using $^{90}$Sr-$^{90}$Y beta source, which emits $\bar{{\nu}_e}$ ($E_{\nu max}$ = 2.28 MeV) in the geoneutrino energy range.

\section{Neutrino Directions}

We consider $\bar{{\nu}_e}$ detection in $\rm(CH_2)_n$ liquid scintillator with a density of $\rm 0.8\; g/cm^3$.
Geoneutrinos detection signature is delayed coincidence between the prompt positron signal and signal from the neutron
capture gammas. Positron energy spectrum boosted by absorption of two 0.511 MeV annihilation quanta is shown in
 Fig.\ 1. For each event positron and neutron capture coordinates (${\bf R}_{ei}$ and ${\bf R}_{ni}$) are reconstructed using
scintillation light collection pattern and (or) distribution of light arriving times detected in photomultipliers.

Information on neutrino incoming direction(s) is obtained from studies of the reconstructed positron-neutron position vectors ${\bf R}_{eni}$: 

\begin{equation}
{\bf R}_{eni} = {\bf R}_{ei} - {\bf R}_{ni} 
\end{equation}

Below we present a number of Monte Carlo experiments, which include modeling of Geoneutrinos coming from mantle, from continental crust and from both of these sources at a time. To check our Monte Carlo simulations we start with geoneutrino parallel beam coming from a remote point-like source already considered at CHOOZ for the reactor  $\bar{{\nu}_e}$'s [8]. For reaction\,(1) differential cross section and kinematics we use results presented in [9,10].

\subsection{Parallel geoneutrino beam}

Low energy neutrons emitted in reaction (1) (their energies are in the 1$-$15 keV range) are strongly concentrated around incoming geoneutrino direction, which is assumed as $Z$ axis (Fig.\ 2). In subsequent collisions the memory of this direction is partially conserved and neutrons are displaced from the reaction (1) point in the positive $Z$ direction. Formation of this displacement in the first few collisions can be traced step by step in Fig.\ 3a,b. After first $\sim 8$ collisions the memory is completely lost and neutrons are slowing down and diffuse symmetrically around the displaced center. Slowed down neutrons are captured mostly in hydrogen with mean capture time $\tau \approx 200\ {\rm \mu s}$. The energy independent neutron - $^{12}$C scattering cross section is 4.8 b, neutron - proton scattering cross section is 20 b for neutron energies grater than 1 eV, for lower energies due to hydrogen binding in the molecule the cross section increases gradually and reaches 65 b at thermal energy of 1/40 eV. Mean square radius $< r^2 >^{1/2}$ of the neutron capture distances from the (displaced) center is calculated to be 6.1 cm. The average neutron displacement $d_Z$ in $Z$ direction is found to equal 1.72 cm:

\begin{equation}
d_Z = 1.72\ {\rm cm}, \qquad \qquad \qquad \langle r^2 \rangle ^{1/2} = 6.1\ {\rm cm}
\end{equation}

The main factor, which limits the sensitivity of the method, is the accuracy of neutron - positron position reconstruction procedure. We assume that vector (2) projections are Gaussian-distributed around their mean values:

$$
\langle {\bf R}_{enx} \rangle = 0 \pm \sigma_x/N^{1/2}, \qquad \langle {\bf R}_{eny} \rangle = 0 \pm \sigma_y/N^{1/2}, \qquad \qquad \qquad \quad
$$

\begin{equation}
\langle {\bf R}_{enz} \rangle = 1.72\ {\rm cm} \pm \sigma_z/N^{1/2}, \qquad \sigma_x = \sigma_y = \sigma_z = \sigma
\end{equation}
where {\sl N} is the number of detected reaction (1) events and $\sigma$ is the Gaussian dispersion.

Assuming now $\sigma$ = 20 cm and ${\sl N}$ = 2500 neutrino events (as in the CHOOZ experiment) we find $\langle{\bf R}_{enz}\rangle = 1.7 \pm 0.4$ cm, and thus neutron displacement can be detected at four standard deviation level. In polar coordinates the most probable direction of the average vector $\langle {\bf R}_{eni} \rangle$ is $\theta$ = 0; the uncertainty of this value is $\delta \theta \approx 2^{1/2}\sigma/1.7 \cdot N^{1/2} \approx 18^o$ in agreement with CHOOZ [8].

Another way of presenting results is to build, in polar coordinates, the angular distribution $W({\cos}\,\theta)$ of reconstructed vectors ${\bf R}_{eni}$ around $\bar{{\nu}_e}$ incoming direction, which can be presented in the form:

\begin{equation}
W({\cos}\,\theta) \approx {\rm Const} \cdot (1+k_Z \cdot {\cos}\,\theta)
\end{equation}

In the case considered here $k_Z \sim (d_Z/\sigma)$. (In Eqs. (5) a higher order term $a \cdot {\cos^2}\theta$ with $a \sim (d_Z/\sigma)^2$ is neglected). We generate $10^5 $ $\bar{{\nu}_e}$ events and find "true" angular distribution: $W({\cos}\,\theta) = {\rm Const} \cdot (1+k_Z \cdot {\cos}\,\theta)$ , with $k_Z = 0.138$ (solid line in Fig.\ 4). Next we generate 2500 events and find the uncertainty $\delta _{k_Z}= \pm 0.03$, represented by shaded area in Fig.\ 4. Thus the angular distribution (5) differs from isotropy at four standard deviation level. We conclude that both ways of analysis (displacement and angular distribution) are equivalent.

\subsection{Geoneutrinos from the Lower Mantle}

Under the crust is the Earth's mantle, which is 2900 km thick. Its lower part (the lower mantle) is believed to contain as much U and Th as the Earth's crust [6,7]. Geoneutrino flux has finite vertical component and is assumed to be symmetric azimuthally (Fig.\ 5). Again we generate $10^5$ Monte Carlo events and find "true" value of neutron cloud displacement $d_{LM} = 1.20$ cm in the vertical direction. Thus

$$
\langle {\bf R}_{enx} \rangle = \langle {\bf R}_{eny} \rangle = 0 \pm \sigma/N^{1/2}, \qquad \qquad \qquad \qquad \qquad \qquad \qquad \quad
$$

\begin{equation}
d_{\rm LM} = \langle {\bf R}_{enz} \rangle = 1.20 \ {\rm cm} \pm \sigma/N^{1/2}.
\end{equation}

With 4000 detected neutrino events $d_{\rm LM} = 1.2 \pm 0.32$ cm can be found.

\subsection{Geoneutrinos from the Crust}

Here we consider hypothetical case where continental crust source forms a uniform 6000 km diameter and 40 km thick circular region centered around the geoneutrino detector. In this model average horizontal components of reconstructed positron-neutron vector ${\bf R}_{en}$ vanish, the vertical component is small:

$$
\langle {\bf R}_{enx} \rangle = \langle {\bf R}_{eny} \rangle = \pm \sigma/N^{1/2}, \qquad \qquad \qquad \qquad \qquad \qquad \qquad \qquad
$$

\begin{equation}
d_{\rm Cr} = \langle {\bf R}_{enz} \rangle \approx 0.29 \ {\rm cm}\pm \sigma/N^{1/2}.
\end{equation}

The displacement for crust geoneutrinos $d_{\rm Cr}$ is 4 times smaller than that for the neutrinos from the lower mantle and cannot be obsereved with the number of detected geoneutrinos $N$ = 4000 considered here.

\subsection{Geoneutrinos from the Crust and the Lower Mantle$\qquad$}

We consider detector installed at a continental site (e.g. at Baksan Observatory). The average displacement of neutron cloud in the vertical direction is given by the expression:

\begin{equation}
d_{\rm LM+Cr} = \langle {\bf R}_{enz} \rangle = \alpha_{\rm LM}d_{\rm LM} + (1- \alpha_{\rm LM}) \cdot d_{\rm Cr}\pm \sigma/N^{1/2},
\end{equation}
where $d_{\rm LM}$ = 1.2 cm, $d_{\rm Cr}$ = 0.29 cm and $\alpha_{\rm LM} = F_{\rm LM}/(F_{\rm LM}+F_{\rm Cr})$ is the lower mantle fraction in the total geoneutrino incoming flux; $N$ = 4000, the maximal achievable events sample considered here, $\sigma/N^{1/2}$ = 0.32 cm.

Dependence (8) is shown in Fig.\ 6. One can see that separation method considered here is not very sensitive. Only sufficiently large displacements $d_{\rm LM+Cr}$, larger than $\approx1$ cm if found, can indicate contradiction to the predictions of the orthodox Earth's model. In case experiment favors lower displacements, and thus indicates low contribution of the mantle geoneutrino flux, the dominant role of the crust geoneutrinos predicted by the model can roughly be confirmed. Only with much larger number of collected events ($N \sim 2\cdot10^4$) and therefore with much larger detector more definite conclusions could be reached.

\subsection{Detector Calibrations}

Detection of small displacements discussed above requires adequate calibration procedures. While usual method, based on inserting neutron and gamma sources into the fiducial volume can and should be exploited, use of sufficiently strong $\bar{{\nu}_e}$ source is highly desirable. We propose to consider for calibration purposes movable $\sim$1\ MCi\  $^{90}$Sr-$^{90}$Y beta source. $^{90}$Sr ($T_{1/2}$ = 28.6 yr) decays to the ground state of $^{90}$Y($E_{\nu max}$ = 2.28 MeV, $T_{1/2}$ = 64 h), which with 99.99\% probability populates the ground state of the stable $^{90}$Zr nucleus. If installed at the distance of 30 m from the 30 kton detector center, the source can generate about $2\cdot 10^5$ events of reaction (1) per year [11]. 

Two circumstances make this source attractive. First, it will irradiate the detector with $\bar{{\nu}_e}$  flux of known intensity, known energy spectrum in the geoneutrino energy range and of known angular structure and, second, the sources are produced commercially and used to supply heat for Radioisotope Thermoelectric Generators (RTGs) (see Fig.\ 7).

We note that proposed calibration method could also be used in other low energy $\bar{{\nu}_e}$ experiments employing large liquid scintillation detectors.

\section{Discussions and Conclusion}
 
In this paper we present a first attempt to analyze directions of a multidirectional low energy $\bar{{\nu}_e}$ flux. In this attempt only two geoneutrino sources have been taken into account: the continental crust and lower mantle. We haven't analyzed perturbations due to possible fluctuations of U and Th concentrations in the detector's immediate vicinity. Clearly more work is needed to come to more accurate results.

At this preliminary stage of analysis we can summarize the results as following:
\begin{itemize}
\item Present understanding of the total amount of radiogenic sources and their distribution in the Earth's reservoirs is based on a shaky ground of cosmogenical and geochemical arguments with an obvious deficit of direct experimental evidence [1, 6, 7, see also discussion in Ref.\ [2]]. Information obtained with one 30-kton target mass detector using directional separation of incoming geoneutrino flux is useful but limited: it can give only some indications against the orthodox Earth's model predictions, or provide it's rough confirmation. More definite information could be obtained only with much larger detector. To avoid misunderstanding we note that the Georeactor hypothesis can be tested at BNO with $\sim$1-kton antineutrino detector [13].
\item With 30-kton scintillation spectrometer additional information on geoneutrinos can be obtained: The total geoneutrino flux can be measured within $\sim$2\% accuracy; accurate spectroscopy of positron spectrum (see Fig.\ 1) will provide important information on global U to Th concentration ratio [12]. ``Geoneutrino" is a part of future studies at BNO of low energy $\bar{{\nu}_e}$ fluxes of natural origin. Among other physical targets we mention: Galactic Supernova explosion and estimation of frequency of gravitational collapses in the Universe by detection of isotropic flux of Relict antineutrino (for more points see Ref.\ [2] and project LENA [14]).
  \end{itemize}
 
\section*{Acknowledgments }

This work is supported by RFBR and RF President's grant 1246.2003.2

\end{document}